# When Effective Field Theories Fail

**John F. Donoghue**
*Department of Physics*
*University of Massachusetts, Amherst*
*Amherst, MA 01002 USA*
*E-mail:* `donoghue@physics.umass.edu`

In this talk, I describe and defend four non-standard claims about four effective field theories, and try to extract some lessons about the limits of effective field theory. The four theses (and a capsule diagnosis given in parentheses) are: 1) Kaon loops are not a reliable part of chiral perturbation theory (dimensional regularization does not know about the chiral scale), 2) Regge physics is inappropriately missing from SCET (an infinite set of scales are needed) 3) There is likely a barrier in the use of EFT in general relativity in the extreme infrared (curvature effects build up) and 4) Gauge non-invariant operators should be included in describing physics beyond the Standard Model (as they could probe the idea of emergent gauge symmetry and falsify string theory).

# 1. Introduction

The title of this talk is admittedly a little provocative. Effective field theories don't themselves fail. However we are not infallible and the theories are sometimes subtle. Since we advertise these theories as being completely rigorous, it is important that we examine what we do in order to to see if it meets the standards of full rigor. In this talk, I will make four claims that challenge our present practice, and I will briefly give supporting arguments. While these describe specific issues, I have tried to frame them in terms of lessons which may be more general. Hopefully a discussion of these claims will be an opportunity to understand our effective field theories better.

## 2. There is no warning light on the machine

My thesis of this section is that kaon loops are not a reliable part of chiral perturbation theory. The diagnosis points to dimensional regularization as the culprit for why we do not readily know this. There can be related issues in other EFTs bucause of the ubiquity of use of dimensional regularization.

There are two key foundations of effective field theory that deal with the separation of heavy degrees of freedom from the light ones in an effective field theory. One is the decoupling theorem (the Appelquist Carrazone theorem [1]) that tell us that the effects of heavy particles go into local terms in a field theory, either renormalizeable couplings or in non-renomalizeable effective interactions suppressed by powers of the heavy mass. The other is the work of Wilson[2] who taught us how to separate the degrees of freedom above and below a given energy scale and then to integrate out all the high energy effects and form a full field theory with the remaining degrees of freedom below the separation scale.

In chiral perturbation theory the $\rho(770)$ is not present in the effective theory so that it is clear that it has been integrated out. Chiral perturbation theory therefore has a separation of the heavy DOF from the light ones at around the rho mass. Indeed, the energy expansion of the tree level Lagrangians readily reveals this mass as the primary scale of chiral perturbation theory. Loop diagrams however run over all energies, in principle even those beyond the scale of the EFT. If we actually used a Wilsonian cutoff separating light from heavy scales, then loop diagrams would reveal this scale also. However, cutoffs are awkward to deal with, so we use dimensional regularization. When regularized dimensionally, loop integrals don't have any sense of where the separation scale of the EFT lies and therefore can include scales beyond it. Indeed the dominant contributions could be from beyond it, for all we know. Moreover, the $m^2 \ln m^2$ chiral logarithms grow without bound and, taken seriously, would eventually violate the decoupling theorem. It is in this sense that there is no warning light - the loop calculation itself does not tell us if it is reliable or not.

It is relatively easy to see that for mesons of mass m, the relevant energy scale in loop diagrams is $4m^2$ rather than $m^2$ - this factor of four will be important. One way to see this is that the right scale is to note that all the kinematic functions that come from loop diagrams are functions of q²/4m². A typical example is the finite part of the standard chiral loop integral which has the form the prediction of the pion electromagnetic form factor [3,4] which is

$$G_\pi(q^2) = 1 + \frac{2L_9^r}{F^2}q^2 - \frac{1}{96\pi^2 F_\pi^2}\left[\left((q^2 - 4m_\pi^2)\, H(\frac{q^2}{m_\pi^2}) + q^2 \ln\frac{m_\pi^2}{\mu^2} + \frac{q^2}{3}\right) + \frac{1}{2}(\pi \to K)\right]$$

where the kinematic parts of the loop is contained in the function

$$\begin{aligned}
H(\frac{q^2}{m^2}) &= 2 + \sqrt{1 - \frac{4m^2}{q^2}}\left[\ln\left|\frac{\sqrt{1 - \frac{4m^2}{q^2}} - 1}{\sqrt{1 - \frac{4m^2}{q^2}} + 1}\right| + i\pi\theta(q^2 - 4m^2)\right] & |q^2| > 4m^2,\ q^2 < 0 \\
&= 2 - 2\sqrt{\frac{4m^2}{q^2} - 1}\ \mathrm{ctn}^{-1}\sqrt{\frac{4m^2}{q^2} - 1} & 0 < q^2 < 4m^2 \\
&= \frac{q^2}{4m^2} & |q^2| \ll 4m^2
\end{aligned}$$

Here we can see the factor of 4m² explicitely.

More generally, we know that we can reconstruct Feynman diagrams from their cuts and poles. For chiral loops, there are only cuts and these start at 4m². For these cuts to reproduce one-loop chiral Feynman diagram, the tree amplitudes due to the cuts must be those predicted by lowest order chiral Lagrangains. The only way that the loops can be reliably predicted is if the trees are themselves reliable at the energy 4m² and beyond.

Let us show this in a rigorously calculable example. There are two chiral sum-rules [5] that express physical examples in terms of known on-shell intermediate states. These state that

$$\begin{aligned}
-4\bar{L}_{10} &= \int_{4m_\pi^2}^\infty \frac{ds}{s}[\rho_V(s) - \rho_A(s)] \\
F_\pi^2 &= \int_{4m_\pi^2}^\infty ds[\rho_V(s) - \rho_A(s)]
\end{aligned}$$

Here $\rho_{V-A}(s)$ is a spectral function that describes the on-shell intermediate states which can be produced by vector or axial-vector currents. It can be extracted from tau decays and from e⁺e⁻ interactions. The full form of this spectral function [6] is shown in Fig 1, with ALEPH data being the dominant contribtuion. The peaks seen are the vector ρ(770) resonance and the axial-vector $a_1(1260)$ resonance.

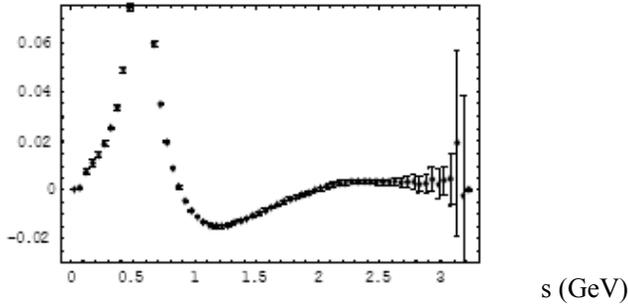

s (GeV)

Fig. 1. The experimental spectral function $\rho_{V-A}(s)$ which enters the dispersive predictions of the chiral parameters.

The quantities being predicted here are the pion decay constant and one of the low energy parameters of the chiral Lagrangian which enters into radiative pion decay, $L_{10}$ [3]. Both of the quantities contain chiral logarithms from loop diagrams. Explicitly

$$F_\pi = F_0[1 - \frac{m_\pi^2}{16\pi^2 F_0^2}\ln\frac{m_\pi^2}{\mu^2} - \frac{m_K^2}{32\pi^2 F_0^2}\ln\frac{m_K^2}{\mu^2} + \frac{4m_\pi^2}{F_0^2}L_5^r(\mu) + \frac{8m_K^2 + 4m_\pi^2}{F_0^2}L_4^r(\mu)]$$

and

$$-4\bar{L}_{10} = 4L_{10}^r(\mu) - \frac{1}{48\pi^2}[\ln\frac{m_\pi^2}{\mu^2} + \frac{1}{2}\ln\frac{m_K^2}{\mu^2} + \frac{3}{2}]$$

The goal is to understand under what conditions the chiral logarithms are reliable.

We can make chiral predictions for the low energy porition of these spectral functions, and this allows us to use these to see how chiral logs arise. The low energy behavior of the spectral function is just a product of tree amplitudes for the p-wave vector-pi-pi coupling and has the form

$$\rho(s) = \frac{1}{48\pi^2}\left[1 - \frac{4m_\pi^2}{s}\right]^{3/2}$$

A blow-up of the very low energy region of the spectral function is shown in Fig. 2, along with the chiral prediction.

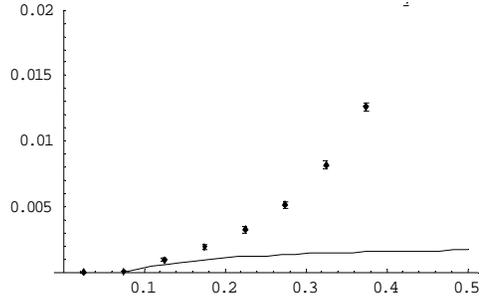

Fig. 2. A blow up of the very low energy end of the spectral function. The solid curve is the chiral prediction, and the points are the data.

We see that the chiral prediction holds at low energies, followed by the ρ and $a_1$ resonances. The ρ and $a_1$ portions (both short distance in nature) go into determining the low energy constants, as is well known in the chiral literature. These sum-rules then allow us to explicitely see the low energy contributions and the short distance physics in a case where we have a good handle on both.

The low energy behavior generates chiral logs. This can be seen by explicit computation. First let us break the integral up into the threshold prediction of chiral symmetry and the higher energy portion, for which we would use the experimental data to get the short distance contibution:

$$-4\bar{L}_{10} = \int_{4m_\pi^2}^{\Lambda^2} \frac{ds}{s} \frac{1}{48\pi^2}(1 - \frac{4m_\pi^2}{s})^{1/2} + \int_{\Lambda^2}^{\infty} \frac{ds}{s}(\rho_V - \rho_A)$$

In this case, the threshold gives the full chiral log contribution to the sum rule:

$$\int_{4m_\pi^2}^{\Lambda^2} \frac{ds}{s} \frac{1}{48\pi^2}(1 - \frac{4m_\pi^2}{s})^{1/2} = -\frac{1}{48\pi^2} \ln \frac{m_\pi^2}{\Lambda^2} + \ldots$$

Likewise for $F_\pi$, we get a chiral log from the threshold region:

$$F_\pi^2 = \int_{4m_\pi^2}^{\Lambda^2} ds \frac{1}{48\pi^2}(1 - \frac{4m_\pi^2}{s})^{1/2} + \int_{\Lambda^2}^{\infty} ds(\rho_V - \rho_A)$$

$$\int_{4m_\pi^2}^{\Lambda^2} ds \frac{1}{48\pi^2}(1 - \frac{4m_\pi^2}{s})^{1/2} = \frac{m_\pi^2}{8\pi^2} \ln m_\pi^2 + \ldots$$

So we see that in the dispersive treatment, the low energy representation of the cut pion tree amplitudes has turned into chiral logs that we normally get from loop integrals. In the case of the pion, these are reliable because the chiral prediction for the vector-pi-pi amplitude is valid at these energies.

What about the kaon? This is a very different situation. It is clear that the kaon threshold starts at 4 $m_K^2 \sim$ 1GeV$^2$. This is well above the onset of short distance physics and we have no expectation that the chiral prediction should be accurate at this energy. Indeed, experimental data on the kaon contribution to the spectral function exists [7], and it does not look at all like the chiral form, see Fig. 3. The chiral prediction for the kaon spectral function is not reliable because the kaon threshold occurs beyond the limit of validity of chiral perturbation theory.

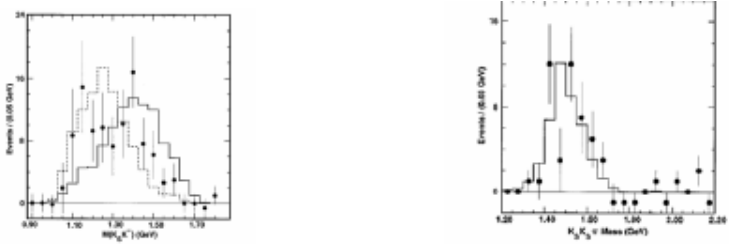

Fig. 3. The kaon contributions to the vector (left) and axial-vector (right) spectral functions.

We have seen that theory and experiment both tell us that kaon loops are not a reliable part of chiral perturbation theory because the relevant expansion parameter $4m_K^2/m_\rho^2$ is larger than unity. This would have been evident if we had used a Wilsonian procedure of including only the physics below $m_\rho$ with a strict separation, i.e. by employing some form of a cutoff on the loops (or the dispersion integral). However, we used dimensional regularization and it returned the $m_K^2 \ln m_K^2$ form without realizing that it came from scales that are firmly outside of the validity of the effective field theory.

One might object that we have had two decades of phenomenology of chiral physics involving kaon loops and the theory works, especially in the meson sector. However, it is almost a theorem that kaon loops are hidden by the chiral low energy constants (LECs) (such as $L_{10}$, or $L_{4,5}$ above). To see this, set the pion mass to zero as this does not change the kaon loops. The kinematic effect depending on energies, such as given above for the pion form factor, do get suppressed as factors of $q^2/m_K^2$. The residual chiral logs turn into factors of $\ln m_K^2/\mu^2$, where $\mu$ is the scale associated with dimensional rgurlarization. The $\ln \mu$ dependence is always associated with a chiral LEC as it is related to the running of the LECs. Therefore we always find the $\ln m_K^2$ also associated with a LEC. At fixed values of $m_K$ then the kaon chiral logs are then hidden by the presence of a chiral LEC and can be accommodated in phenomenology by a shift in the LEC.[1]

An exception to this statement is in the chiral extrapolation business. If you change the kaon mass, the chiral logs give a special non-analytic form that is different from the analytic behavior of the LECs. The chiral logs give a strong curvature to a chiral extrapolation in the meson mass. However, it is now clear that the lattice data do not show the effects of chiral logs for masses around the kaon mass. (See J. Flynn's review at this conference [8] or L. Lellouch's review at Lattice 2008 [9].) My argument above explains why this is the case - chiral perturbation theory should not contain these loops for mesons of this mass. The use of cutoff regularization has also been used to argue this point [10], but I have made the argument without any use of models. Indeed we see that the quantitative criteria for reliability of the loop diagrams is $4m^2 \ll m_\rho^2$ or $m \ll 385$ MeV.

The other exception to the statement about hiding the effects of kaon loops is in the baryon sector. There are non-analytic terms that are not logarithmic, i.e. $m^3$. For example in the baryon masses there are very large $m_K^3$ effects. These are not readily hidden by a low energy constant, and do have a bad effect on the phenomenology, creating corrections that are 100% of the size of the leading term. By going to high orders in the chiral expansion and adding very large counterterms, one is able to adjust the final results to match the physical values [11]. However the chiral expansion breaks down in the process, becoming perilously close to the form $M = M_0 [1 - 1 + 1 - ...]$. Similar excessively large effects occur elsewhere in baryon chiral perturbation theory with kaons. Moreover, in addition to these phenomenological problems, there are theoretical problems [12]. Analysis of these loop diagrams shows that almost all of the dimensionally regularized result for the loop integral comes from short distance physics. This is shown in Fig. 4, from [12] where the long distance portion, calculated with a cutoff regularization with a cutoff $\Lambda$, is seen to be small and obey decoupling, while the dimensionally regularized form violates decoupling badly as the $m^3$ terms grow rapidly without bound . While the $m^3$ non-analytic behavior does appear in both results at very small meson masses, it is no longer reliable at the physical value of the kaon mass and the residual effects below the cutoff are much smoother. Again this is corroborated in lattice results (displayed by M. Savage at this conference [13]) which are quite smooth without evidence of the strong chiral non-analytic behavior.

---

[1] The caveat that keeps this from being a real theorem is that there could be further analytic term proportional to $m_K^2$ without the log, which emerges distinct from the LEC. However, I know of no such "naked" effects in standard phenomenology

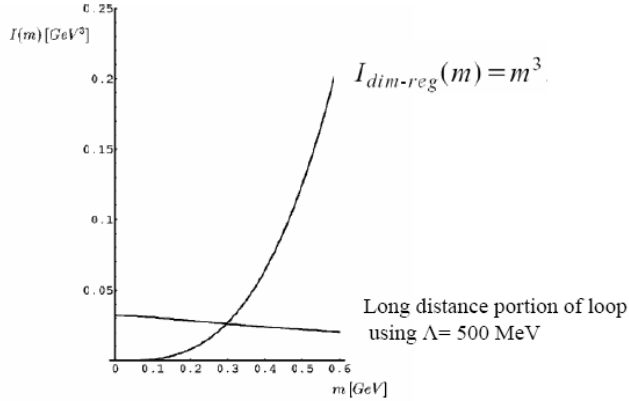

Fig. 4. The result for the Feynman integral giving the non-analytic $m^3$ dependence using dimensional regularization and also a cutoff regularization.

I have argued on theoretical grounds that kaon loops are not a reliable part of chiral perturbation theory, and pointed to phenomenology which supports this claim. The diagnosis points to the use of dimensional regularization, which has no scale, as the reason why this was not evident previously. This is one of the subtleties that come when we think of effective field theory in terms of energy scales but employ a loop regulator that does not know about these scales.

**3**. **Way too many scales**

The separation of different scales is one of the cornerstones of effective field theory. But what would happen if a given process required effectively an infinite hierarchy of scales? I will argue that this is required if Regge behavior is to be incorporated in the Soft Collinear Effective Theory (SCET), and is perhaps one of the reasons that this has not yet been accomplished. Moreover, I will argue that many of the results of SCET cannot be viewed as firmly established until the Regge issue is fully understood.

Daniel Wyler and I are in the process of putting out a paper describing this in detail [14], so I will concentrate on the underlying logic here and refer the interested reader to that paper for the formulas.

Regge theory describes physics in the high energy kinematic region $s \to$ infinity with momentum transfer t fixed. The key feature for our discussion here is that with the exchange of Reggeons, the amplitudes become functions of $s^{\alpha(t)}$ where $\alpha(t) = \alpha(0) + \alpha' t$ is the Regge exponent. The leading high energy behavior of scattering amplitudes is given by "soft" Pomeron exchange with $\alpha(0) = 1.02$. In cases where t is fixed but large enough, Lipatov and collaborators [15] have shown perturbatively that QCD exhibits Regge behavior, with a Reggeized gluon and a "hard" Pomeron. From both this theoretical work and from experimental efforts over many years, it is clear that Regge physics is part of QCD.

On the other hand, SCET has been developed to describe the interactions of high energy quarks and gluons through the exchange of soft and collinear particles [16]. This includes the Regge region, so why has Regge behavior not been understood in SCET yet? This issue is relevant to the applications of SCET. For example, BBNS [17] first argued that certain B decay amplitudes factorize and that soft final state interactions phases are of order $1/m_B^2$ - this is supported in SCET. However, if the Regge behavior can enhance the final state rescattering behavior by a power of $s^{\alpha(0)}$, with $s \sim m_B^2$ and $\alpha(0) = 1.02$, the suppression of soft phases is completely removed. This was argued to be the case in Ref. [18] and remains possible due to the lack of understanding of Regge in SCET. Other SCET statements are also suspect due to $s^{\alpha(0)}$ power corrections until Regge behavior is understood within the theory.

The origin of Regge behavior in field theory is understood. We refer the reader in particular to the book by Ross and Forshaw [19] which provides a thorough treatment. Regge behaviour comes from multi-loop ladder diagrams describing particle exchanges in the t-channel, as in Fig 5. It is the kinematics that is relevant for the comparison with SCET.

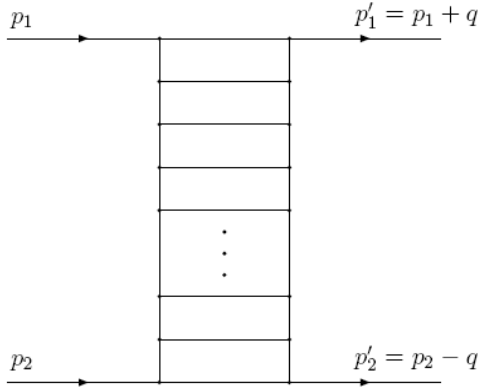

Fig. 5. The ladder diagrams whose sum gives Regge behaviour in field theory.

Analysis of a general n-rung ladder diagram shows that the energy denominator, after Feynman parameterization, behaves as

$$[x_1 x_2 ... x_n s + D(x_i, y_i, t)]^n$$

In general this diagram would fall as $s^{-n}$ except near the kinematic corner with all $x_i = 0$. In this corner, one obtains a logarithmic result

$$\frac{g^2}{s(n-1)!}[\beta(t) \ln(-s)]^{n-1}$$

When the infinite set of ladder diagrams are summed, the logs are promoted to a power $s^{\beta-1} = s^{\alpha}$. It is this feature of logs turning into powers which can upset the present predictions of SCET.

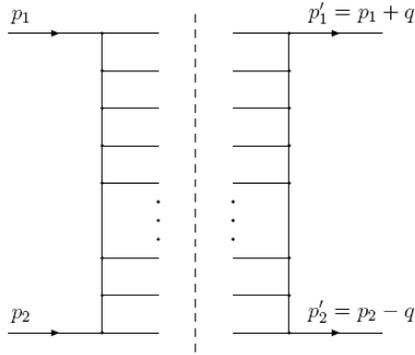

Fig. 6. The cut ladder diagrams. The rungs of the ladder are all on-shell.

In order to see what kinematics this corner of parameter space corresponds to, it is best to reconstruct the loop diagrams from the cuts - in this case we need only the s-channel cut that crosses the n- rungs of an n-loop ladder diagram, see Figure 6. For the cut, the gluons of the rungs are on-shell and the legs of the ladder are off shell by an amount of order $t^{1/2}$. Let us describe this in the center-of-mass with the top particle moving to the right with energy $s^{1/2}$ and the bottom particle moving to the left with the same energy. In SCET variables the right moving and left moving light cone variables are

$$n^\mu = (1,0,0,1), \ \bar{n}^\mu = (1,0,0,-1)$$

respectively. As one moves down the rungs of the ladder, the momentum on a vertical leg is constrained by the requirement that the cut rung be on-shell. In words, the on-shell condition for the rungs of the ladder requires that the momenta be ordered in a special way. In particular, as one goes down from the top of the diagram, each of the legs has to carry less "right-moving" momenta than the one above it, and also more "left-moving momenta" than the one above it. So the legs are ordered with decreasing "rightness" going down, and increasing "leftness", interpolating between the right-moving top line and the left-moving bottom line. This part is purely kinematics.

However, it is also possible to show that the factors of ln(s) only arise when the legs are **strongly ordered** - i.e. with the step down in "rightness" being parametrically large. Let us describe this decrease in momentum by a multiplicative factor η treating it as a small parameter. If the leg has almost the same momentum as the one above it, the logarithms do not occur and Regge behaviour is not obtained. This strong ordering results in each of the rungs being directionally collinear with the one above it, but with a parametrically smaller energy.

This can be described in formulas, but the sense is best conveyed by an illustration. Let us imagine the ten loop diagram at the energy $10^5$ GeV. If we use η =1/10, the momenta would be

$$
\begin{aligned}
p_1 &= 10^5 n \\
k_1 &= 10^4 n + 10^{-5} \bar{n} + k_\perp \\
k_2 &= 10^3 n + 10^{-4} \bar{n} + k_\perp \\
k_3 &= 10^2 n + 10^{-3} \bar{n} + k_\perp \\
k_4 &= 10^1 n + 10^{-2} \bar{n} + k_\perp \\
k_5 &= 10^0 n + 10^{-1} \bar{n} + k_\perp \\
k_6 &= 10^{-1} n + 10^0 \bar{n} + k_\perp \\
k_7 &= 10^{-2} n + 10^1 \bar{n} + k_\perp \\
k_8 &= 10^{-3} n + 10^2 \bar{n} + k_\perp \\
k_9 &= 10^{-4} n + 10^3 \bar{n} + k_\perp \\
k_{10} &= 10^{-5} n + 10^4 \bar{n} + k_\perp \\
p_2 &= 10^5 \bar{n}
\end{aligned}
$$

What we see here is a series of legs that are mostly collinear to the one next to it in the chain, but with rightness decreasing as one goes down and leftness increasing. There is a soft (really Glauber) particle in the middle of the chain. In reality, η is not a fixed factor but is integrated over. However, the ln (s) behavior does not come from the region where η ~1, so it is not a normal collinear gluon of SCET. In fact, for η <<1, the rung is collinear to the leg above it but is parametrically smaller in energy – we can call this an ordered gluon. It appears different than the other gluons in the SCET classification, although perhaps it can be accounted for as the kinematic endpoint of a collinear gluon. There is a soft particle (technically a Glauber particle since the transverse momentum is larger than either longitudinal momentum) somewhere down this chain, where dominant right-movers communicate with dominant left-movers.

For a SCET description we would define a series of parametrically smaller collinear particles. One can't readily obtain the right kinematics with only a single collinear type of gluon, because all the action would occur at the kinematic endpoint of the domain of validity. Since one sums loops from one to infinity, this would in principle need to be done for an infinite number of times as s is infinitely large. This is a challenge for effective field theories. In practice, Regge behavior seems to turn on in QCD at a several GeV, so that approximate treatments with finite s appears to be sufficient. However, the multiple scales seem to be an intrinsic feature of the Regge kinematics.

## 4. Making a mountain out of a molehill

The usual criterion for the utility of effective field theory is that the field configurations carry little energy, or equivalently have little space-time variation in coordinate space. But even if the fields are slowly varying, their effect can build up if they continue over long enough distances. I will argue that an effect of this nature happens in the application of effective field theory to gravity[2]. I am unsure of the importance of the effect, but nevertheless it is an interesting non-standard feature that deserves to be understood better.

---

[2] This is a refinement of an argument given in [20]

General relativity displays the characteristics of an effective field theory [20]. The curvature carries two derivatives of the fields and (ignoring the small cosmological constant) the action can be expanded in a derivative expansion. Around flat space the usual momentum behavior is obtained

$$V(q^2) = \frac{GMm}{q^2}\left[1 + a'G(M+m)\sqrt{-q^2} + b'G\hbar q^2 \ln(-q^2) + c'Gq^2\right]$$

and if one Fourier transforms to coordinate space one obtains (in the same order)

$$V(r) = -\frac{GMm}{r}\left[1 + a\frac{G(M+m)}{rc^2} + b\frac{G\hbar}{r^2c^3}\right] + cG^2Mm\delta^3(r)$$

describing a long distance expansion. Explicit calculation leads to a parameter free prediction for the quantum correction due to general relativity [21]

$$V(r) = -\frac{GMm}{r}\left[1 + 3\frac{G(M+m)}{rc^2} + \frac{41}{10\pi}\frac{G\hbar}{r^2c^3}\right]$$

However there are indications that at very long distances, this nice expected behavior could be spoiled. For example consider a horizon. The horizon itself is not the problem locally. If we are in free fall through a horizon of a large distant black hole, we could still construct a locally flat set of coordinates and use effective field theory in this neighborhood if the curvature is small enough. However, in this situation we could not extend our coordinates out to infinity - no signal can propagate to spatial infinity. So a potential such as described above is not correct at all distances - failing in the extreme infrared. Even if the quantum effect is extremely tiny in this region, as an issue of principle the power dependence must be modified.

In addition, the Hawking Penrose theorems tell us that most space-times contain curvature singularities somewhere in the past, present or future. If we have a region that locally has a small curvature but nevertheless try to extend our effective field theory to a distance space time point (also with small curvature) on the other side of a singularity, the effective theory fails to describe the intermediate region.

These hand-waving arguments suggest that it is not just the local curvature that plays a role but the cumulative effect of the curvature integrated over long distances. The diagnosis can be made more concrete by considering Reimann normal coordinates [23]. At any point we can define a set of coordinates that look locally Minkowski, with the deviations being described by an expansion in the curvature

$$g_{\mu\nu}(y) = \eta_{\mu\nu} + \frac{1}{3}R_{\mu\alpha\nu\beta}(y_0)y^\alpha y^\beta - \frac{1}{6}R_{\mu\alpha\nu\beta;\gamma}(y_0)y^\alpha y^\beta y^\gamma$$
$$+ \left[\frac{1}{20}R_{\mu\alpha\nu\beta;\gamma\delta}(y_0) + \frac{2}{45}R_{\alpha\mu\beta\lambda}(y_0)R^\lambda_{\gamma\nu\delta}(y_0)\right]y^\alpha y^\beta y^\gamma y^\delta + \mathcal{O}(\partial^5)$$

Here we see that in a neighborhood of the origin we are roughly flat and well-behaved. For small curvatures we can use effective field theory techniques in this region. However, for any R there always is a distance where R$y^2$ becomes of order unity. These coordinates are not good to use perturbatively any more. We could go to that distant point and again use a set of Reimann coordinates in that neighborhood. But in either case, we give up the description connecting the two distant regions[3]. So not only does R have to be small but so does some integrated value of R through distance. The criteria for the effective field theory seems to be not only $G/r^2 \ll 1$ (a parameter that defines the UV limit of the theory), but also we seem to need $R_{\mu\nu\alpha\beta}\, r^2 \ll 1$ (which is a parameter which describes the IR limits of the local expansion).

This idea of an IR expansion parameter representing in some sense the integrated curvature explains how horizons and singularities can challenge the techniques of effective field theory. Curvature effects built up over long distances eventually result in barriers to local calculations. This happens also to the classical corrections when treated perturbatively, but we know how to solve the classical equations exactly, leading to the recognition of horizons, etc. For the quantum corrections, we have only the perturbative result and are not sure how to proceed non-perturbatively. In principle these effects could happen in other theories also, for example with photons or if the chiral fields are treated as massless. This is a novel phenomenon in effective field theory and seems outside of our present techniques.

## 5. Who'd 'a thunk it?[4]

Finally let's discuss the case where an effective field theory treatment is inadequate because a seemingly innocuous assumption turns out to be inappropriate. There are probably a quasi-infinite number of ways to illustrate this option, but I will choose one that interests me at the moment. I will argue that it is potentially a way to differentiate string theory from emergent gauge theories.

We use effective field theory to look for physics beyond the Standard Model. The rules of the game are to write all operators consistent with the gauge symmetries of the Standard Model, to order them by dimension and use the lowest dimensional ones in phenomenology [24]. I will argue that emergent gauge theories (should they really exist) would lead to operators outside of this classification - specifically to gauge-non-invariant operators.

The search for physics beyond the Standard Model is dominated by unification theories. In these, the gauge symmetries that we see at low energy not only persist to high energy, but are enhanced by being unified into higher gauge groups with larger symmetries[5]. This is of course an intriguing idea. String theory represents a particularly striking completion of the unification paradigm; the many new degrees of freedom are tied together into a finite and highly symmetric framework.

---

[3] This example used the metric, but the Hawking Penrose theorem says that a similar effect happens in the curvature.

[4] My father used to use this phrase when something particularly surprised him. Apparently it is traceable to a dummy - Mortimer J. Snerd - and plays on the correspondence that drink/drunk should imply think/thunk.

[5] If I am trying to be provocative, I may as well argue that we have no historical precedent for the idea of gauge unification. Maxwell's feat was not the unification of two gauge theories, but rather the identification of the single correct gauge theory. Likewise, the electroweak theory is also not gauge unification but rather gauge mixing. The two gauge groups involved do not get united, although the gauge bosons mix. So gauge unification would be a novel phenomena if it occurred. To be fair there is also no historical precedent for emergent gauge symmetries, although there are plenty precedents for emergent theories in general.

There is also a different possibility that has been explored far less fully. This is the idea that the gauge symmetries could be low energy emergent phenomena, coming from a fundamental theory without gauge degrees of freedom. If the underlying theory had a finite number of degrees of freedom, this could be a finite theory. This phenomenon occurs in certain condensed matter systems [25]. The original Hamiltonian does not have a U(1) (or higher) gauge symmetry, but the ground state of the theory does manifest this and contains gauge bosons. Holgar Nielsen [26] has an attractive argument that gauge theories may emerge from an underlying random dynamics. The idea of emergent symmetry faces significant hurdles and may turn out to not be possible for the fundamental interactions. However at present it remains an intriguing possibility.

If the gauge symmetry is not fundamental, there could be residual gauge violating interactions. For example, when one integrates over momentum in a loop integral one encounters photon states at the highest energies. We know that if we use a cutoff regularization we can violate gauge invariance. In this case, the lack of fundamental photons at the highest energies would appear like a cutoff, and since the theory does not have the gauge invariance at these energies, there would be good reason to expect a violation of gauge invariance. This would not happen in the unification paradigm since the symmetry persists to all energy. If general covariance is also emergent, we would expect a violation of it also. Indeed, gravity might be the best place to look for anomalous effects because gravitational interactions are already suppressed by the Planck scale.

These considerations argue for allowing operators into the general effective Lagrangian which violate gauge symmetries and general covariance. The standard rules of the game have been formulated with the unification paradigm in mind, but this could be incorrect. We have started a study of such effects [27]. A potentially important outcome is that a gauge-violating signal would serve to falsify stung theory. In string theory gauge invariance is exact at high energies and we would not expect signals of gauge violation.

## 6. Concluding comments

Effective field theory is a set of powerful techniques which allow us to isolate the appropriate degrees of freedom and use these in making predictions. The use of EFT has become common, but it is still subtle in many instances. This talk explores some of the limits (in energy, in multiplicity of scales, in integrated effects and in hidden assumptions) which could upset our present practices in effective field theory. I hope that a discussion of these phenomena can lead to a better understanding of effective feld theory in general.